\documentclass[aps,prl,twocolumn]{revtex4}
\usepackage{graphics}
\usepackage{graphicx}

\begin{document}

\bibliographystyle{apsrev}

\title{Accurate determination of the scattering length of metastable Helium atoms using dark resonances between atoms and exotic molecules}

\author{S. Moal, M. Portier, J. Kim, J. Dugu\'{e}, U. D. Rapol, M. Leduc, and C. Cohen-Tannoudji}


\affiliation{Ecole Normale Sup\'{e}rieure and Coll\`{e}ge de France, \\
Laboratoire Kastler Brossel, 24 rue Lhomond, 75231 Paris Cedex 05,
France.}

\date{Received: date / Revised version: date}

\begin{abstract}

We present a new measurement of the s-wave scattering length $a$ of spin-polarized helium atoms in the $2^3S_1$ metastable state. Using two-photon photoassociation spectroscopy and dark resonances we measure the energy $E_{v=14}=-91.35\pm0.06$~MHz of the least bound state $v=14$ in the interaction potential of the two atoms. We deduce a value of $a = 7.512 \pm 0.005$~nm, which is at least one hundred times more precise than the best previous determinations and is in disagreement with some of them. This experiment also demonstrates the possibility to create exotic molecules binding two metastable atoms with a lifetime of the order of $1~\mathrm{\mu s}$. 

\end{abstract}

\maketitle

The s-wave scattering length, which characterizes the interactions between ultracold atoms, is essential to describe the structure and the dynamics of Bose Einstein Condensates (BEC). After BEC was achieved for metastable helium atoms in the $2^3S_1$ state \cite{Robert,Pereira2001}, several estimates for their s-wave scattering length $a$ were derived (see Fig.1), which lack accuracy and are not all in good agreement with each other. We present in this letter a new determination of $a$, at least one hundred times more precise than the best previous ones, solving unambiguously all the discrepancies in the literature. This important constant is critical to interprete previous or future experiments dealing with the hydrodynamical regime in the ultracold gas \cite{Leduc}, the metastable helium BEC in an optical lattice, or $^4He$-$^3He$ mixtures \cite{Stas}. To measure $a$, we use two-photon photoassociation (PA) and dark resonances to perform a spectroscopic measurement of the binding energy $E_{b_2}$  of the least bound molecular state $b_2$ ($v=14$) in the  interaction potential $^5\Sigma_g^+$ between spin-polarized metastable helium atoms. The precise value of $E_{b_2}$ which we obtain can be also used to check the predictions of ab initio recent quantum chemistry calculations of the  interaction potential, which can be very accurate for an atom as simple as  $^4He$ having only two electrons and no hyperfine structure. Another interesting point is that the two-photon process which we use prepares an exotic molecular state binding two atoms with a very high internal energy of about 20 eV. From the line shape analysis of our spectra, we are also able to estimate the lifetime of this exotic state to about 1~$\mu s$. This information could stimulate theoretical models trying to understand the effect of Penning processes between two metastable atoms interacting in a bound state rather than in a free collisional state.

The two-photon scheme is sketched in Fig.~2: it involves a pair of colliding atoms in a state 0 with energy $E_\infty$ and two molecular bound states $b_1$ and $b_2$; here $b_1$ is the $v=0$ level in the purely long range $0_u^+$ potential previously studied by our group \cite{Leonardexp}. Such $\Lambda$-type excitation schemes were used with alkali atoms to realize the spectroscopy of molecules in the electronic ground state in ultracold gases \cite{Tolra,Lisdat,Schloder}, to find scattering lengths \cite{Weiner} and very recently to observe sub-natural linewidth quantum interference features \cite{Dumke}. In some other experiments molecules in the electronic ground state have  been coherently formed from a BEC~\cite{Wynar,Winkler}. Oscillations between atomic and molecular quantum gases have recently been observed~\cite{Ryu}. The results obtained here with metastable $^4He$ atoms are simpler to interprete theoretically because of the absence of hyperfine stucture, which reduces the number of collision channels involved in the light-assisted collisional process. Here, we do not use a BEC of metastable $^4He$ atoms in order to avoid mean-field shifts. Finally, we use ultracold temperatures such that the dispersion $k_BT$ of scattering energies $E_\infty$ is very small compared to the widths $\hbar \gamma_1$ and $\hbar \gamma_2$. As a consequence, our signal is not affected by thermal averaging \cite{Napolitano94}.

\begin{figure}[ht]
\begin{center}
\resizebox{0.9\columnwidth}{!}{
\includegraphics[scale=0.35]{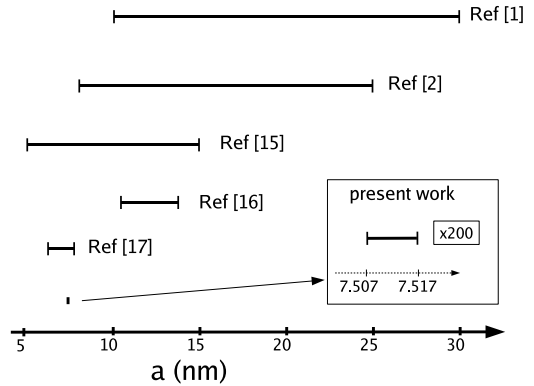}
}
\end{center}
\caption{Comparison of the present result with previous experimental determinations of the s-wave scattering length $a$ of spin-polarized metastable helium: from the expansion of the condensate \cite{Robert,Pereira2001}, from the evaporative cooling rate \cite{Tol}, from the observation of inelastic collisions \cite{Seidelin} and from light-induced frequency-shifts in one-photon photoassociation \cite{Kim2005}.}
\label{fig:values-of-a}
\end{figure}

\begin{figure}[ht]
\begin{center}
 \resizebox{0.7\columnwidth}{!}{
\includegraphics[scale=0.8]{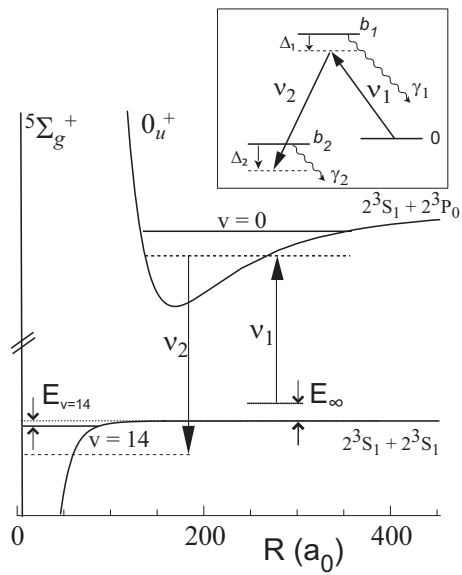}
}
\end{center}
\caption{Levels involved in the two-photon photoassociation experiment. $^5\Sigma_g^+$ is the interaction potential between two spin-polarized $2^3S_1$ helium atoms, $v=14$ is the least bound state in this potential. Laser $L_1$ at frequency $\nu_1$ operates on the free-bound transition with a detuning $\Delta_1$, laser $L_2$ at frequency $\nu_2$ drives the bound-bound transition with a detuning $\Delta_2$ from the two-photon transition. The levels $b_1$ ($v=0$) and $b_2$ ($v=14$) are characterized by their decay rates $\gamma_1$ and $\gamma_2$.} \label{fig:potentiel}
\end{figure}

The metastable helium sample is confined in a magnetic trap of three-coil Ioff\'e-Pritchard type with a bias field of 3~G and operated slightly above condensation at a temperature of a few~$\mu$K and at a density of a few 10$^{13}~$~cm$^{-3}$. The PA beams of lasers $L_1$ and $L_2$ are derived from a DBR diode laser operating in an external cavity (linewidth of order 600~kHz) at 1083 nm. Its frequency is tuned close to the 0-$b_1$ transition, 1.4~GHz red of the $2^3S_1$-$2^3P_0$ atomic transition. The frequencies $\nu_1$ and $\nu_2$ are generated from this laser using acousto-optical modulators (AOM), which ensures that both laser beams are coherent and eliminates the frequency jitter of the lasers.  The two superimposed laser beams are  focused through the same optical fiber (waist size of 300~$\mu m$) on the atomic cloud which is much smaller (a few tens of $\mu m$). The pulse duration is adjusted to optimize the PA signals. We registered signals with a fixed value of the frequency $\nu_2$ of laser $L_2$ and used laser $L_1$ as a probe, scanning its frequency $\nu_1$ in the vicinity of the free-bound transition 0-$b_1$. We measured the temperature raise of the cloud, as well as the decrease of optical density and the loss of atoms. The three methods give the same resonance positions, but the temperature raise signal has the best quality \cite{Leonardexp}.

In a first series of experiments, we used large intensities $I_2$ such that laser $L_2$ induces a coupling (Rabi frequency $\Omega$) between $b_1$ and $b_2$ larger than the linewidth $\gamma_1/2\pi=3$MHz of the excited state $b_1$ ($\Omega \gtrsim \gamma_1$). In Fig.~3a the frequency $\nu_2$ is set close to resonance with the $b_1$-$b_2$ transition ($\Delta_1 \approx \Delta_2$). The Rabi frequency is $\Omega/2\pi=4$~MHz. We clearly identify a double-peak structure corresponding to an Autler-Townes splitting~\cite{Autler}: the two molecular bound states $b_1$ and $b_2$ are dressed with the light field of laser $L_2$, forming a doublet probed by laser $L_1$. We checked that the separation between the two peaks increases linearly with the square root of the intensity $I_2$ according to the Rabi frequency $\Omega$. We also studied the case when laser $L_2$ is set farther from resonance with the $b_1$-$b_2$ transition ($\Delta_1 \neq \Delta_2$). One of the two peaks gets closer to the frequency of the 0-$b_1$ transition while the second one, shown in Fig.~3b, becomes asymmetric and is associated with a stimulated Raman process between 0 and $b_2$. When increasing $\Delta_1-\Delta_2$ one progressively goes from the purely Autler-Townes case to the stimulated Raman effect. In the case where $\nu_2$ is set such that the two peaks are symmetrical, the abscissa of the middle point between the two peaks in Fig. 3a can be used to deduce the energy of the $b_1$-$b_2$ transition, from which the energy $E_{b_2}$ can be derived. An extrapolation to zero laser intensity is necessary to take light shifts into account. 

In order to derive a very precise measurement of $E_{b_2}$ we performed another series of experiments at smaller intensity $I_2$ ($\Omega \lesssim \gamma_1$) taking advantage of the phase coherence between the two lasers in our set-up. Here, the intensity $I_2$ is not high enough to induce an Autler-Townes splitting. We find dark resonances occurring when the Raman resonance condition $E_\infty+h\nu_1-h\nu_2=E_{b_2}$ is satisfied. They are interpreted as an interference feature: a linear superposition of states 0 and $b_2$ is created such that a destructive interference occurs between the absorption amplitude of photon 1 on the $0$-$b_1$ transition and of photon 2 on the $b_1$-$b_2$ transition. The photoassociation process is thus  blocked and a very narrow dip appears at the centre of the main PA line when laser $L_1$ is scanned (see Fig. 3c). The dark resonance observed in our experiment does not result in a complete suppression of the PA signal, due to the finite lifetime of level $b_2$. This method offers several advantages. The position of the dark resonance is independent of the position of the excited state $b_1$ and is not affected by a Zeeman shift, as both $0$ and $b_2$ states are expected to have nearly the same magnetic moment~\footnote{In the absence of hyperfine coupling in $^4He$, the only interaction which mixes different spin states and could induce a different magnetic moment for the states 0 and $b_2$ is the very weak spin-dipole interaction.}. The three solid lines of Fig.~3 represent a fit based on the theory of \cite{BJ1996,BJ1999} and taking only into account the heating by the decay product of $b_1$. The detailed analysis of the asymmetric lineshape of Fig.~3b, typical of a Fano profile (see \cite{Lounis} and ref. therein), shows that the heating due to $b_2$, which would give rise to a symmetrical profile with a similar width, is negligible (see Fig.~4a of \cite{Koelemeij}). We have also checked that a small but non negligible contribution of $b_2$ to the heating would not change the center of the symmetrical dark resonance of Fig.~3c used to measure $E_{v=14}$.
\begin{figure}[ht]
\begin{center}
\resizebox{0.9\columnwidth}{!}{
\includegraphics[scale=0.7]{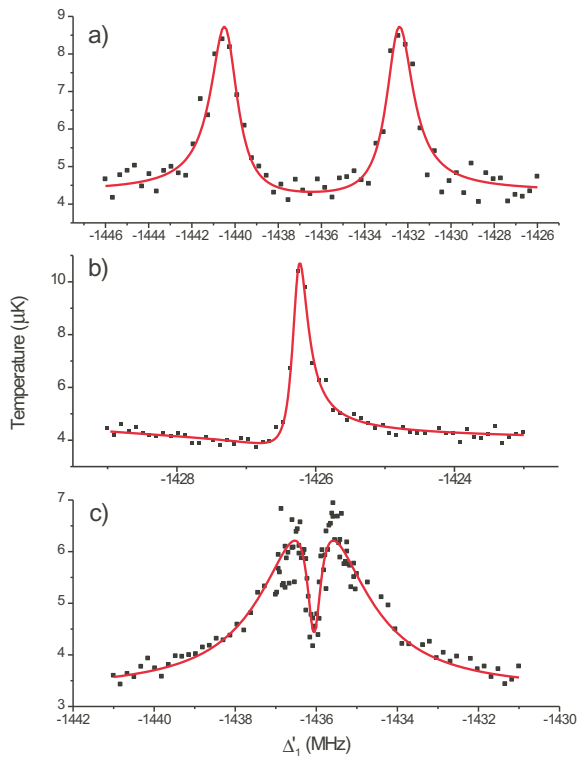}
}
\end{center}
\caption{Two-photon PA experiments with metastable helium. Temperature of the cloud as a function of frequency detuning $\Delta'_1$ of laser $L_1$ from the atomic resonance for a fixed frequency $\nu_2$ of laser $L_2$ (detuning from the atomic resonance $\Delta'_2$= -1345.0 MHz, -1335.0 MHz, -1344.6 MHz in Fig.~3a, 3b and 3c respectively). Laser intensities: $I_1=$7  mW.$\mathrm{cm^{-2}}$   $I_2$=330 mW.$\mathrm{cm^{-2}}$ in Fig.3a, $I_1=140$ mW.$\mathrm{cm^{-2}}$   $I_2=33$ mW.$\mathrm{cm^{-2}}$ in Fig.~3b  and  $I_1=$7 mW.$\mathrm{cm^{-2}}$   $I_2=7$ mW.$\mathrm{cm^{-2}}$ in Fig.~3c. Data are remarkably well fitted using the theory of references \cite{BJ1996,BJ1999} (see text).  Fig.~3a shows the Autler Townes doublet, Fig 3b the Fano profile of the Raman signal for $\Delta_1-\Delta_2\approx$ 10 MHz and  Fig.~3c a narrow dip attributed to a dark resonance signal. The Zeeman shift of the one-photon transition and the temperature shift have to be added to the value of $\Delta'_1$ to compare it with the binding energy $E_{v=0}=-1418.1$~MHz quoted in \cite{Kim2004}. Note the differences between the horizontal scales in the three figures.} \label{fig:parabola}
\end{figure}

Several dark resonance spectra were recorded for atomic clouds prepared with a temperature between 2 and 10~$\mu$K while keeping the atomic density in the range of a few $\mathrm{10^{13}~cm^{-3}}$. The temperature $T$ is evaluated with an estimated precision of 20\% by fitting the cloud expansion in free flight. Since $k_BT$ is small enough ($k_BT\lesssim \hbar\gamma_2$), the calculation of the thermal averaging of the lineshape shows that the width and shape of the resonance do not depend on $T$ but its position is shifted by an amount equal to $3k_BT/2$. This prediction has been checked experimentally: Fig.~4 shows a linear decrease of the resonance parameter $\nu_1$-$\nu_2$ as a function of $T$, with a slope which matches $3k_B/2h$ within 15\%. For the very low intensities used here, we have checked that various corrections associated with light shifts, mean field interaction and laser fequency measurement are smaller that the error bar given for our result. We found no Zeeman shift of the two-photon transition when varying the magnetic field between 0.5 and 6 G. For the extrapolated position at zero temperature we thus adopt the value and the conservative error bar indicated on the vertical axis in Fig.~4. Finally, we find the value $E_{v=14}=-91.35\pm0.06~\mathrm{MHz}$ for the binding energy of level $b_2$~($v=14$) .

From this measured value, we derive the s-wave scattering length of spin-polarized metastable helium. We use the most precise available interaction potentials derived by Przybytek and Jeziorski from ab initio calculations. In reference \cite{Przybytek}, a Born-Oppenheimer potential is provided, as well as its upper and lower bounds estimated from the extrapolated calculation to a complete molecular basis set. We use several potentials within these bounds by scaling the inner part of the original Born-Oppenheimer potential. For each potential the energy of the least-bound state $E_{v=14}$ and the scattering-length $a$ can be calculated, which yields a relation between $E_{v=14}$ and $a$. We check that the adiabatic, relativistic and quantum electrodynamic corrections and their uncertainties as calculated in \cite{Przybytek} introduce errors on the determination of $a$ smaller than those due to the experimental uncertainty on $E_{v=14}$. To be conservative, the error bar on $a$ obtained from the experimental measurement is therefore just doubled in our final result: $a=7.512\pm0.005$~nm.

We also derive an estimate of the lifetime of the $v=14$ molecular state by studying the width of the Raman peak (see Fig.~2b) which is directly related to $\gamma_2$ for large detunings $\Delta_2-\Delta_1$. We additionnaly confirm this result using the fit of the dark resonance lineshape of Fig.~3c. The temperature and laser intensity that we use are checked to have no influence on the width. We thus find $0.05~\mathrm{MHz}<\gamma_2/2\pi<0.3~\mathrm{MHz}$, corresponding to an intrinsic lifetime between 0.5 and 3~$\mu s$ for the molecule. The main causes of inelastic decay for this molecule are likely to be spin relaxation and relaxation induced ionization \cite{Fedichevspin,Venturiclosecoupled}, giving a roughly estimated lifetime of 4 $\mu$s quoted in \cite{Koelemeij}, which is in good agreement with the presently measured linewidths. However, other decay mechanisms like atom-molecule collisions could occur. We plan to add ion detection to our setup to get more insight on the decay mechanism of $b_2$ since it has been shown that no ions are produced from the decay of $b_1$.

Our results can be compared to some very recent experiments on atom-molecule dark states. In a Bose-Einstein condensate of Rb \cite{Winkler}, the width of the dark resonance was due to atom-molecule collisions in the limit of vanishing laser intensity. In an ultracold sample of Na atoms the linewidth was limited by the temperature \cite{Dumke}. In our specific case, the linewidth is not temperature limited although we are not in the condensate regime, because the ground molecular-state $b_2$ is very short-lived compared to the alkali case.
\begin{figure}[th]
\begin{center}
 \resizebox{0.9\columnwidth}{!}{
\includegraphics[scale=1]{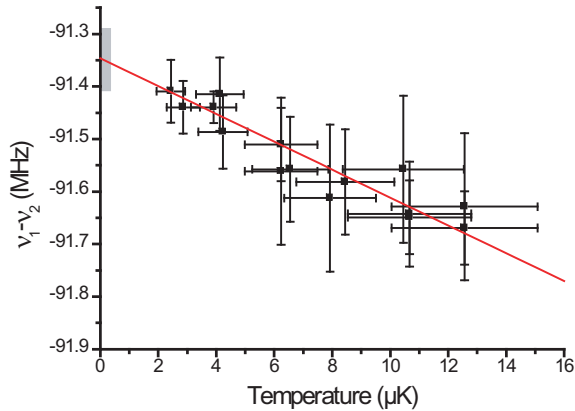}
}
\end{center}
\caption{Position $\nu_1-\nu_2$ of the dark resonance signal (see Fig 3c), as a function of temperature $T$. The slope of the straight line is compatible with a dependence of $3k_BT/2$ in unit of $h$ (see text). On the vertical axis we indicate the final error bar which we attribute to the binding energy of the molecule in the $v=14$ state.}    \label{fig:temperature}
\end{figure}

In conclusion, we prepared exotic, doubly excited molecules in the very weakly bound state $v=14$ in the interaction potential between two spin-polarized metastable helium atoms. We measured its binding energy $E_{v=14}=-91.35\pm 0.06~\mathrm{MHz}$ providing a very stringent experimental test of ab initio calculations \cite{Starckmeyer,Gadea2004,Przybytek}. The most precise theoretical value $E_{v=14}=-87.4\pm 6.7$~MHz which has been recently reported~\cite{Przybytek} is in very good agreement with our measurement. We deduced the value of the s-wave scattering length between two metastable atoms $a=7.512 \pm 0.005$~nm, which is by far the most accurate experimental determination yet reported. We find very good agreement with our previous determination ($a=7.2\pm0.6$~nm) based on light-induced frequency-shifts in PA experiments. The discrepancy with the value found in \cite{Seidelin} ($a=11.3 \: ^{+2.5}_{-1.0}$~nm) remains to be explained. Finally, we gave an estimation of the lifetime of the exotic molecular state of the order of 1 $\mu$s, which could be useful for the theory of Penning ionization.

The authors want to thank Allard Mosk, J\'er\'emie L\'eonard and Matthew Walhout for their contribution at an early stage of this work, as well as for their constant interest and suggestions.

\end{document}